\documentclass[]{mn2e}
\usepackage{color,soul}
\usepackage{graphicx}
\usepackage{deluxetable}
\usepackage{lscape}
\usepackage{epsfig}
\usepackage{amssymb,amsmath}
\newcommand{\OVI}{\mbox{O\,{\sc vi}}}
\newcommand{\HI}{\mbox{H\,{\sc i}}}
\newcommand{\MgII}{\mbox{Mg\,{\sc ii}}}
\newcommand{\MgI}{\mbox{Mg\,{\sc i}}}
\newcommand{\FeII}{\mbox{Fe\,{\sc ii}}}

\newcommand {\apgt} {\ {\raise-.5ex\hbox{$\buildrel>\over\sim$}}\ }
\newcommand {\aplt} {\ {\raise-.5ex\hbox{$\buildrel<\over\sim$}}\ }

\title[New Low-Redshift DLAs]{ Thirty-six New, High-Probability, Damped Ly$\alpha$ Absorbers at  Redshift $0.42<z<0.70$}

\author[D. A. Turnshek et al.]
{\parbox[t]{\textwidth}{\raggedright 
David A. Turnshek$^{1}$\thanks{E-mail: turnshek@pitt.edu},
Eric M. Monier$^{2}$,
Sandhya M. Rao$^{1}$,
Timothy S. Hamilton$^{3}$, 
Gendith M. Sardane$^{1}$, and
Ryan Held$^{2}$
}
\vspace*{6pt}\\
$^{1}$Department of Physics and
Astronomy and PITTsburgh Particle physics, Astrophysics, and Cosmology Center 
(PITT PACC),\\
 University of Pittsburgh, Pittsburgh, PA 15260, USA\\
$^{2}$Department of Physics, The College at Brockport, State University
of New York, Brockport, NY 14420, USA\\
$^{3}$Department of Natural Sciences, Shawnee State University, Portsmouth, Ohio 45662, USA}

\begin{document}

\date{}

\pagerange{\pageref{firstpage}--\pageref{lastpage}} \pubyear{2013}

\maketitle

\label{firstpage}

\begin{abstract}
Quasar damped Ly$\alpha$ (DLA) absorption line systems with redshifts $z<1.65$ are used to trace neutral gas over approximately 70\% of the most recent history of the Universe. However, such systems fall in the UV and are rarely found in blind UV spectroscopic surveys. Therefore, it has been difficult to compile a moderate-sized sample of UV DLAs in any narrow cosmic time interval.  However, DLAs are easy to identify in low-resolution spectra because they have large absorption rest equivalent widths. We have performed an efficient strong-\MgII-selected survey for UV DLAs at redshifts $z=[0.42,0.70]$ using HST's low-resolution ACS-HRC-PR200L prism. This redshift interval covers $\sim1.8$ Gyr in cosmic time, i.e., $t\approx[7.2,9.0]$ Gyrs after the Big Bang. A total of 96 strong \MgII\ absorption-line systems identified in SDSS spectra were successfully observed with the prism at the predicted UV wavelengths of Ly$\alpha$ absorption. We found that 35 of the 96 systems had a significant probability of being DLAs. One additional observed system could be a very high $N_{\rm HI}$ DLA ($N_{\rm HI} \sim 2\times10^{22}$ atoms cm$^{-2}$ or possibly higher), but since very high $N_{\rm HI}$ systems are extremely rare, it would be unusual for this system to be a DLA given the size of our sample. Here we present information on our prism sample, including our best estimates of $N_{\rm HI}$ and errors for the 36 systems fitted with damped Ly$\alpha$ profiles. This list is valuable for future follow-up studies of low-redshift DLAs in a small redshift interval, although such work would clearly benefit from improved UV spectroscopy to more accurately determine their neutral hydrogen column densities. 

\end{abstract}


\section{Introduction}

Since the first spectroscopic survey for intervening quasar damped Ly$\alpha$ (DLA) absorption-line systems (Wolfe et al. 1986), it has been recognized that these gaseous regions with neutral hydrogen column densities $N_{\rm HI} \ge 2 \times 10^{20}$ atoms cm$^{-2}$ can be used to trace the neutral gas component of the Universe.  DLA and related observations allow the Universe to be probed over about 90 per cent of its current age.\footnote{Throughout we assume a cosmology with H$_0$ = 70 km s$^{-1}$ Mpc$^{-1}$, $\Omega_{\rm M} = 0.3$, and $\Omega_{\Lambda} = 0.7$.} The most recent extensive results from low-redshift UV and high-redshift optical DLA surveys have been presented by Rao, Turnshek, \& Nestor (2006; hereafter RTN2006) and Noterdaeme et al. (2012), respectively. Relative to RTN2006 (41 DLAs in their statistical sample), Meiring et al. (2011) found a very high incidence of DLAs in a more recent HST COS UV blind survey (3 DLAs), which is counter to what Bahcall et al. (1993) found in the HST FOS UV QSO Absorption-Line Key Project blind survey (1 DLA), but both UV blind surveys suffer from small number statistics. Most recently, Noterdaeme et al. (2014) present a study of $\sim 100$ extremely strong high-redshift DLAs, with $N_{\rm HI} \ge 5 \times 10^{21}$ atoms cm$^{-2}$.
Aside from statistical results on the DLA incidence (product of absorber cross-section and their comoving number density) and the cosmic mass density of neutral gas, many interesting facets of galaxy formation and evolution can be considered through follow-up studies of DLA metallicities, dust, molecular fractions, star formation, kinematics, associated galaxies, and clustering as a function of redshift. These topics in DLA research have been widely discussed in the literature, but it is not the purpose of this paper to summarize them (see Wolfe et al. 2005 for a past review).    

Here, we simply wish to emphasize that while follow-up studies of DLAs are providing a wealth of information about the neutral-gas-phase component of the Universe, reasonably accurate statistical measurements in any relatively narrow low-redshift interval have remained elusive. By low redshift we mean redshifts $z<1.65$, for which the Ly$\alpha$ line falls in the UV.  
For example, the low-redshift DLA survey of RTN2006 probed $z<1.65$. This redshift regime corresponds to $\sim$  70 per cent ($\sim 9.6$ Gyrs) of the most recent history of the Universe, where significant evolution is known to have occurred.  But since observation of a low-redshift Ly$\alpha$ line requires space-based UV spectroscopy, the number of confirmed DLAs remains relatively small in all narrow low-redshift intervals (e.g., in cosmic time intervals corresponding to  $\sim 1$ to $2$ Gyrs).  Even with the efficient strong-\MgII-selected DLA survey method used by RTN2006, only 9 DLAs have been previously confirmed at $z=[0.42, 0.70]$. This redshift interval spans $\sim1.8$ Gyr in cosmic time, $t\approx[7.2,9.0]$ Gyrs after the Big Bang.

A number of years ago, when STIS had failed and COS was not yet installed on HST, we had the opportunity to perform a large strong-\MgII-selected survey for DLAs at $z=[0.42,0.70]$. This is a very efficient way to conduct a DLA survey since any biases can be accounted for if the statistics of \MgII\ absorption-line systems are well understood (RTN2006).  This redshift interval was chosen because it is well-matched to the sensitivity of HST's low-resolution ACS-HRC-PR200L prism. To demonstrate this, Figure 1 shows the Ly$\alpha$ damping profiles for a range of DLA $N_{\rm HI}$ values within this redshift interval at full resolution and at the resolution of a prism spectrum. The prism has nonlinear dispersion so that a prism spectrum of a Ly$\alpha$ line at $z=0.42$ has resolution $\lambda$/$\Delta\lambda \approx 197$ (4.4 \AA\ pixel$^{-1}$ at 1726 \AA), while at $z=0.70$ it has $\lambda$/$\Delta\lambda \approx 97$ (10.6 \AA\ pixel$^{-1}$ at 2067 \AA). With this in mind, we undertook prism observations of the UV Ly$\alpha$ absorption lines in 109 strong-\MgII\ systems to determine if they exhibited an absorption profile with damping wings. The data collection on these 109 systems required 105 single-orbit observations since four of the quasars had two systems.\footnote{Note that during our program there were three one-orbit observations with target-acquisition failures. Repeat single-orbit observations were obtained for two of the failed observations; the observation of J003740.13-090810.0 was not repeated. Thus we lost one of the original strong-\MgII\ systems in our sample and it does not appear in our lists.}

Of the 109 strong-\MgII\ systems in our sample that were observed with the prism, we found that only 97 of them had observations that were usable in our DLA search. We also obtained spectroscopy of three quasars with known DLAs (``DLA calibrators'') to confirm our ability to recognize and measure DLAs. In general, if a Ly$\alpha$ line does not exhibit a significant damping profile (i.e., if $N_{\rm HI}$ is too low in a \MgII\ system), it would be undetectable or nearly undetectable in its prism spectrum. Thus, prism observations offer a suitable method for distinguishing DLAs from non-DLAs. 

\begin{figure}
\vspace{-0.4in}\centerline{
\includegraphics[width=1.0\columnwidth,clip,angle=0]{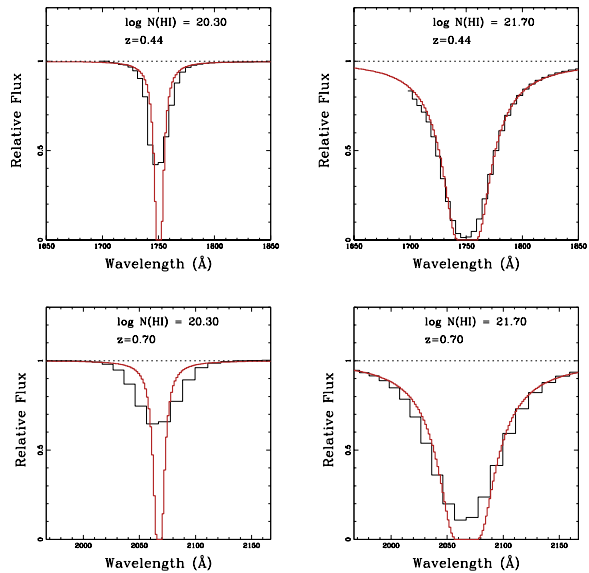}}
\caption{Simulation of expected HST ACS-HRC-PR200L prism spectra of DLAs at $z=0.44$ and $z=0.70$ for a low $N_{\rm HI}$ value of $2\times10^{20}$ atoms cm$^{-2}$ and a high $N_{\rm HI}$ value of $5\times10^{21}$ atoms cm$^{-2}$. The smooth red profiles (see online version) are the normalized theoretical DLA absorption-line profiles. The black profiles are the expected spectra at the resolution of the prism. Note the poorer resolution at higher redshift. Also, the redshift range of our usable observations is taken to be $z=[0.42,0.70]$, however at $z<0.44$ (see the top right panel) the short wavelength side of a high-$N_{\rm HI}$ DLA will begin to become impossible to measure reliably over the expected range in $N_{\rm HI}$ values since the prism calibration does not extend below $\sim 1700$ \AA.}
\label{fig:spectrum}
\end{figure}

Below we describe the results of our HST observations with the primary goal of identifying new DLAs ($N_{\rm HI} \ge 2\times10^{20}$ atoms cm$^{-2}$) at $z=[0.42,0.70]$. Our method to achieve this was simply to examine the prism spectra at the predicted locations of the Ly$\alpha$ absorption lines in strong-\MgII\ absorption-line systems.\footnote{Strong-\MgII\ absorption lines usually means ones with observed \MgII$\lambda 2796$ rest equivalent widths $\ge 0.3$ \AA, however, the systems we study here generally have rest equivalent widths $>1$ \AA.}

The outline of this paper is as follows. In \S2 we present the details of our observational program and data processing for the observed strong-\MgII\ systems. In \S3 we present $N_{\rm HI}$ results for the 36 systems which, based on the prism spectra, have a significant or an interesting probability of being DLAs, along with results on the three DLA calibrators. In \S4 we conclude with a short discussion of the significance of our results.  The implications of these results for measurements of the incidence and cosmic mass density of neutral gas will be discussed in a future paper by Rao et al. 

\section{Sample, Observations, and Data Processing}

The SDSS \MgII\ systems that were chosen for HST prism observations were selected as follows. First, based on preliminary results on the sensitivity of the ACS-HRC-PR200L prism, we decided that it was possible to use it to identify and measure DLAs with $z=[0.37,0.70]$ in the spectra of sufficiently bright quasars.  Second, we used an early version of the Pittsburgh SDSS Quasar \MgII\ Absorption-line Survey Catalog (see Quider et al. 2011) to identify all \MgII\ systems in this redshift interval that met the following quasar brightness and metal-line absorption criteria: (1) quasar SDSS g-band mag brighter than $\sim20$ and (2) \MgII\ rest equivalent width $W^{\lambda2796}_{0} \apgt 1.0$ \AA, which generally (not always) is consistent with having \FeII\ rest equivalent width $W^{\lambda2600}_{0} \apgt 0.3$ \AA\ (but in cases with $z_{abs} \aplt 0.47$ information on $W^{\lambda2600}_{0}$ is unavailable or unreliable in SDSS spectra). Some of the identified quasars also had weaker \MgII\ systems in their spectra, and these appear in Table 1. Additionally, we removed absorption systems from this sample if the predicted location of the DLA would lie too close to broad Ly$\beta$$+$\OVI\ emission in the quasar spectrum (within $\pm 3300$ km s$^{-1}$), since this could cause unnecessary confusion when measuring any DLAs in the low-resolution prism spectra. To achieve our desired final sample size we then decided to limit our observations to include only quasars with g-band magnitudes brighter than $\sim19.25$. Unfortunately, it turned out that the sensitivity of the prism was only sufficient to identify DLAs at $z=[0.42,0.70]$; in the end, five observations of \MgII\ systems with $z<0.42$ were made but were unusable for a DLA search. The findings and methods discussed in RTN2006 can be used with the above selection criteria, and with the results of this study, to arrive at an independent determination of the incidence and cosmic mass density of neutral gas at $z=[0.42,0.70]$.  As noted in the Introduction, this will be quantified in the future paper by Rao et al. The future Rao et al. paper will also incorporate results on other UV surveys for DLAs.   

Table 1 presents the sample of 109 strong-\MgII\ absorbers observed at the location of Ly$\alpha$ with HST's low-resolution ACS-HRC-PR200L prism. Included for each \MgII\ absorber in Table 1 are the SDSS quasar's J2000 name, the total exposure time for the prism observation, the quasar's g-band magnitude, and the quasar's emission redshift; this information is taken from the most recent entries found in the SDSS DR7 and DR9 quasar catalogs of Schneider et al. (2010) and P\^aris et al. (2012), respectively. Also included in Table 1 is information on the metal-line absorption properties for each strong-\MgII\ absorber using results from the online version of the Pittsburgh SDSS Quasar \MgII\ Absorption-line Survey Catalog of Quider et al. (2011). This includes the \MgII\ absorption redshift and metal-line absorption rest equivalent widths and errors for \MgII$\lambda2796$, \MgII$\lambda2803$, \MgI$\lambda2852$, and \FeII$\lambda2600$. Finally, this same information is listed at the end of Table 1 for each of the three quasar ``DLA calibrators'' with known $N_{\rm HI}$ (see Rao et al. 2003 and RTN2006), which were also observed with the prism. We chose to observe these three DLAs because they covered a range of a factor of $\sim7$ in \HI\ column density, and we wanted to assess how well their \HI\ column densities could be re-determined over this range. In Table 1 we adopt the naming convention used in Rao et al. (2003) and RTN2006 for these calibrators but note that Q1629+120 is J163145.17+115603.3, Q2328+0022 is J232820.38+002238.2, and Q2353$-$0028 is J235321.62$-$002840.6. 

\begin{table*}
\caption{The Observed Prism Sample}
\begin{tabular}{lcccccccccc}
\hline
\hline
Quasar  &Exp. time & SDSS g & $z_{em}$ & $z_{abs}$ & \MgII\ $W^{\lambda2796}_{0}$  & \MgII\ $W^{\lambda2803}_{0}$ 
 & \MgI\ $W^{\lambda2852}_{0}$  & \FeII\ $W^{\lambda2600}_{0}$  & Status\tablenotemark{a}\\
 & (s)  & (mag) & & & (\AA)& (\AA) & (\AA)& (\AA)& \\
\hline

J001855.22$-$091351.1 & 2178 & 17.45 & 0.756 & 0.5838 & 1.412$\pm$0.063 & 0.996$\pm$0.061 & 0.015$\pm$0.054 & 0.508$\pm$0.091 & A \\
J013209.75$-$082349.9 & 2178 & 19.25 & 1.121 & 0.6467 & 1.390$\pm$0.098 & 1.409$\pm$0.110 & 0.636$\pm$0.124 & 1.013$\pm$0.099 & A \\
J020046.46+132018.9 & 2186 & 18.86 & 0.883 & 0.6933 & 1.547$\pm$0.073 & 1.440$\pm$0.079 & 0.274$\pm$0.074 & 1.184$\pm$0.094 & B \\
J021044.03$-$082513.3 & 2178 & 18.27 & 1.094 & 0.5458 & 1.131$\pm$0.060 & 0.954$\pm$0.067 & 0.074$\pm$0.061 & 0.584$\pm$0.049 & B \\
J021820.11$-$083259.4 & 2178 & 18.15 & 1.217 & 0.5899 & 3.217$\pm$0.089 & 2.728$\pm$0.088 & 1.337$\pm$0.101 & 2.119$\pm$0.097 & A \\
J031154.53$-$070741.9 & 2178 & 18.73 & 0.633 & 0.4608 & 1.366$\pm$0.076 & 1.052$\pm$0.082 & 0.081$\pm$0.066 & \nodata & B \\
J032253.46$-$071310.7 & 2178 & 18.54 & 0.573 & 0.4320 & 1.972$\pm$0.094 & 2.075$\pm$0.116 & 0.477$\pm$0.105 & \nodata & C \\
J073810.91+401408.8 & 2286 & 18.43 & 1.083 & 0.5799 & 2.247$\pm$0.061 & 1.985$\pm$0.064 & 0.484$\pm$0.064 & 1.347$\pm$0.077 & B \\
J074043.44+344407.7 & 2218 & 18.68 & 0.846 & 0.4783 & 1.671$\pm$0.072 & 1.986$\pm$0.077 & 0.918$\pm$0.069 & 1.744$\pm$0.114 & A \\
J074054.05+332006.5 & 2218 & 17.43 & 0.887 & 0.4917 & 1.252$\pm$0.023 & 1.124$\pm$0.023 & 0.588$\pm$0.024 & 0.941$\pm$0.028 & A \\
J074816.97+422509.3 & 2286 & 17.15 & 1.107 & 0.5579 & 1.520$\pm$0.024 & 1.351$\pm$0.024 & 0.509$\pm$0.031 & 1.093$\pm$0.024 & A \\
J075011.79+222627.6 & 2194 & 18.45 & 0.615 & 0.5816 & 1.039$\pm$0.079 & 1.026$\pm$0.086 & 0.079$\pm$0.059 & 0.614$\pm$0.092 & B \\
J080208.93+360417.7 & 2246 & 18.16 & 1.201 & 0.6789 & 1.145$\pm$0.055 & 0.849$\pm$0.055 & 0.040$\pm$0.050 & 0.394$\pm$0.047 & B \\
J080231.91+381538.5 & 2246 & 18.51 & 0.640 & 0.5670 & 1.065$\pm$0.096 & 0.758$\pm$0.081 & 0.185$\pm$0.095 & 0.435$\pm$0.100 & B \\
J080601.80+275108.2 & 2202 & 18.80 & 0.779 & 0.6922 & 2.101$\pm$0.091 & 1.954$\pm$0.096 & 0.407$\pm$0.079 & 1.012$\pm$0.075 & A \\
J081318.84+501239.8 & 2388 & 18.65 & 0.571 & 0.3829 & 1.519$\pm$0.141 & 1.216$\pm$0.148 & -0.015$\pm$0.106 & \nodata & C \\
J081706.57+261132.1 & 2202 & 18.69 & 0.774 & 0.5234 & 2.194$\pm$0.097 & 1.707$\pm$0.102 & 0.441$\pm$0.127 & 1.111$\pm$0.119 & C \\
J081748.20+493821.4 & 2334 & 18.48 & 1.027 & 0.6855 & 0.987$\pm$0.060 & 0.990$\pm$0.065 & 0.332$\pm$0.061 & 0.622$\pm$0.087 & A \\
J081758.11+302933.8 & 2218 & 18.53 & 1.483 & 0.6943 & 1.429$\pm$0.064 & 0.937$\pm$0.055 & 0.255$\pm$0.072 & 0.763$\pm$0.090 & B \\
J082103.50+400903.3 & 1143 & 18.55 & 1.229 & 0.5994 & 1.167$\pm$0.072 & 0.867$\pm$0.066 & 0.494$\pm$0.070 & 0.611$\pm$0.079 & B \\
J083900.67+370901.4 & 2246 & 18.47 & 0.754 & 0.4697 & 1.168$\pm$0.077 & 0.819$\pm$0.068 & 0.133$\pm$0.077 & \nodata & A \\
J084200.76+333214.8 & 2218 & 18.32 & 0.972 & 0.4750 & 1.703$\pm$0.119 & 1.433$\pm$0.100 & 1.157$\pm$0.131 & 0.643$\pm$0.134 & A \\
J090724.38+493524.0 & 2334 & 18.67 & 0.938 & 0.3953 & 1.553$\pm$0.152 & 1.065$\pm$0.179 & 0.169$\pm$0.121 & \nodata & C \\
J090757.59+421823.6 & 2286 & 18.30 & 0.809 & 0.4540 & 1.453$\pm$0.129 & 0.873$\pm$0.108 & 0.119$\pm$0.085 & \nodata & A \\
J090757.59+421823.6 & \nodata & \nodata & \nodata & 0.5106 & 1.511$\pm$0.089 & 1.376$\pm$0.086 & 0.639$\pm$0.085 & 1.214$\pm$0.102 & A \\
J091158.24+031628.2 & 2172 & 18.86 & 1.177 & 0.5864 & 1.248$\pm$0.108 & 1.306$\pm$0.131 & 0.267$\pm$0.430 & 0.883$\pm$0.118 & B \\
J091750.67+011534.5 & 2172 & 18.97 & 1.011 & 0.5281 & 1.074$\pm$0.142 & 1.140$\pm$0.140 & 0.350$\pm$0.168 & 0.482$\pm$0.142 & B \\
J091759.25+031627.4 & 2172 & 18.16 & 1.056 & 0.6960 & 2.099$\pm$0.066 & 1.454$\pm$0.068 & 0.506$\pm$0.107 & 0.999$\pm$0.074 & B \\
J092026.81+063948.1 & 2178 & 18.20 & 0.821 & 0.4442 & 1.898$\pm$0.116 & 1.809$\pm$0.133 & 0.583$\pm$0.095 & \nodata & B \\
J092845.17+383113.4 & 2246 & 17.80 & 0.718 & 0.6030 & 1.195$\pm$0.064 & 0.892$\pm$0.084 & 0.324$\pm$0.112 & 0.453$\pm$0.075 & B \\
J094735.09+583046.3 & 2452 & 17.80 & 0.935 & 0.5346 & 1.673$\pm$0.057 & 1.463$\pm$0.050 & 0.448$\pm$0.064 & 1.185$\pm$0.064 & B \\
J095648.58+383339.0 & 2246 & 18.76 & 1.195 & 0.6196 & 3.173$\pm$0.128 & 2.422$\pm$0.178 & 0.649$\pm$0.094 & 2.456$\pm$0.090 & A \\
J095740.06+080732.2 & 2178 & 18.76 & 0.870 & 0.6975 & 1.120$\pm$0.072 & 1.169$\pm$0.068 & 0.377$\pm$0.075 & 0.813$\pm$0.085 & A \\
J095844.07+054941.0 & 2178 & 17.85 & 0.730 & 0.6557 & 2.011$\pm$0.052 & 1.976$\pm$0.053 & 0.470$\pm$0.055 & 1.255$\pm$0.048 & A \\
J095848.10+621839.0 & 2466 & 18.65 & 1.237 & 0.5357 & 1.141$\pm$0.068 & 1.086$\pm$0.073 & 0.364$\pm$0.123 & 1.050$\pm$0.105 & A \\
J100154.03+582306.0 & 2452 & 18.70 & 0.837 & 0.6941 & 1.590$\pm$0.072 & 1.289$\pm$0.070 & 0.326$\pm$0.081 & 0.959$\pm$0.090 & B \\
J101211.11+073949.9 & 2178 & 18.28$*$ & 1.030 & 0.6164 & 1.654$\pm$0.103 & 1.556$\pm$0.099 & 0.465$\pm$0.114 & 1.281$\pm$0.107 & A \\
J101911.86+010522.9 & 2172 & 18.63 & 1.137 & 0.6676 & 1.118$\pm$0.067 & 0.821$\pm$0.071 & 0.001$\pm$0.068 & 0.448$\pm$0.058 & B \\
J103219.71+522342.9 & 2388 & 18.77 & 1.198 & 0.6115 & 1.211$\pm$0.106 & 0.895$\pm$0.098 & 0.365$\pm$0.101 & 0.512$\pm$0.092 & B \\
J103733.58+622640.3 & 2466 & 18.52 & 0.907 & 0.5096 & 1.033$\pm$0.092 & 1.040$\pm$0.099 & 0.136$\pm$0.084 & 0.699$\pm$0.142 & B \\
J104146.77+523328.2 & 2388 & 16.63 & 0.678 & 0.4494 & 1.582$\pm$0.038 & 1.276$\pm$0.040 & 0.154$\pm$0.034 & \nodata & B \\
J104754.60+661157.5 & 2550 & 18.40 & 0.471 & 0.4399 & 2.059$\pm$0.106 & 1.230$\pm$0.078 & 0.325$\pm$0.051 & \nodata & C \\
J105823.13+600805.6 & 2466 & 18.44 & 1.071 & 0.6436 & 1.923$\pm$0.125 & 1.552$\pm$0.138 & 0.363$\pm$0.129 & 1.320$\pm$0.142 & A \\
J111013.80+523607.1 & 2388 & 17.77 & 1.004 & 0.5561 & 1.331$\pm$0.049 & 1.266$\pm$0.061 & 0.215$\pm$0.048 & 1.295$\pm$0.061 & A \\
J111456.73+675423.1 & 2550 & 18.70 & 0.941 & 0.6986 & 1.948$\pm$0.066 & 1.786$\pm$0.063 & 0.495$\pm$0.059 & 1.529$\pm$0.077 & B \\
J112016.64+093323.5 & 2178 & 17.85 & 1.099 & 0.4938 & 2.142$\pm$0.095 & 1.616$\pm$0.090 & 0.275$\pm$0.130 & 0.960$\pm$0.105 & B \\
J112319.78+620028.7 & 2466 & 18.38 & 0.662 & 0.5873 & 1.037$\pm$0.089 & 0.907$\pm$0.084 & 0.315$\pm$0.116 & 0.270$\pm$0.080 & B \\
J112357.74+662255.6 & 2550 & 18.59 & 0.834 & 0.4763 & 1.025$\pm$0.072 & 1.121$\pm$0.078 & 0.302$\pm$0.064 & 0.881$\pm$0.096 & B \\
J113245.50+431637.9 & 2314 & 17.34 & 1.027 & 0.5407 & 1.295$\pm$0.042 & 0.997$\pm$0.048 & 0.126$\pm$0.050 & 0.593$\pm$0.054 & B \\
J113823.71+013924.8 & 2172 & 18.71 & 1.042 & 0.6130 & 2.622$\pm$0.098 & 2.315$\pm$0.104 & 0.972$\pm$0.108 & 1.670$\pm$0.095 & A \\
J114059.28+552332.8 & 2452 & 18.01 & 0.643 & 0.5312 & 1.057$\pm$0.074 & 0.885$\pm$0.077 & 0.091$\pm$0.069 & 0.595$\pm$0.104 & B \\
J120200.65+544241.7 & 2388 & 18.63 & 0.835 & 0.4336 & 1.032$\pm$0.108 & 1.060$\pm$0.125 & 0.092$\pm$0.108 & \nodata & C \\
J120449.73+095335.2 & 2178 & 17.97 & 1.276 & 0.6401 & 2.469$\pm$0.122 & 2.557$\pm$0.135 & 0.370$\pm$0.106 & 1.870$\pm$0.127 & A \\
J120611.23+581308.0 & 2172 & 17.05 & 1.192 & 0.6750 & 1.586$\pm$0.026 & 1.201$\pm$0.029 & 0.250$\pm$0.033 & 0.622$\pm$0.030 & B \\
J120743.71+592648.2 & 2452 & 18.42 & 0.989 & 0.5765 & 2.274$\pm$0.085 & 2.022$\pm$0.084 & 0.322$\pm$0.090 & 1.273$\pm$0.087 & A \\
J121753.03+050030.8 & 2178 & 18.52 & 0.632 & 0.5413 & 1.284$\pm$0.071 & 1.689$\pm$0.088 & 0.487$\pm$0.076 & 1.388$\pm$0.104 & A \\
\hline
\end{tabular}
\end{table*}

\begin{table*}
	\addtocounter{table}{-1} \small
\caption{Continued}
\begin{tabular}{lcccccccccc}
\hline
\hline
Quasar  &Exp. time & SDSS g & $z_{em}$ & $z_{abs}$ & \MgII\ $W^{\lambda2796}_{0}$  & \MgII\ $W^{\lambda2803}_{0}$ 
 & \MgI\ $W^{\lambda2852}_{0}$  & \FeII\ $W^{\lambda2600}_{0}$  & Status\tablenotemark{a}\\
 & (s)  & (mag) & & & (\AA)& (\AA) & (\AA)& (\AA)& \\
\hline
J122322.62+553632.2 & 2452 & 18.72 & 1.045 & 0.6773 & 1.465$\pm$0.078 & 0.979$\pm$0.069 & 0.430$\pm$0.071 & 1.075$\pm$0.110 & B \\
J122716.54+091409.2 & 2178 & 18.53 & 0.722 & 0.6979 & 1.213$\pm$0.087 & 1.061$\pm$0.085 & 0.029$\pm$0.072 & 0.317$\pm$0.097 & B \\
J123945.39+531804.9 & 2388 & 18.14 & 1.180 & 0.6942 & 1.692$\pm$0.061 & 1.717$\pm$0.059 & 0.521$\pm$0.075 & 1.420$\pm$0.060 & B \\
J124312.52+480838.4 & 2186 & 17.55 & 0.703 & 0.6053 & 0.268$\pm$0.048 & 0.104$\pm$0.046 & 0.039$\pm$0.045 & 0.121$\pm$0.045 & A \\
J124312.52+480838.4 & \nodata & \nodata & \nodata & 0.6313 & 2.766$\pm$0.053 & 2.113$\pm$0.054 & 0.378$\pm$0.053 & 1.169$\pm$0.060 & B \\
J124512.46+570457.3 & 2452 & 17.85 & 0.951 & 0.5187 & 2.037$\pm$0.090 & 1.624$\pm$0.082 & 0.460$\pm$0.093 & 1.363$\pm$0.106 & B \\
J130544.31+530136.2 & 2388 & 16.88 & 0.864 & 0.5632 & 2.365$\pm$0.035 & 2.131$\pm$0.043 & 0.509$\pm$0.047 & 1.555$\pm$0.043 & B \\
J130544.31+530136.2 & \nodata & \nodata & \nodata & 0.6148 & 0.450$\pm$0.037 & 0.324$\pm$0.035 & 0.060$\pm$0.037 & 0.360$\pm$0.042 & A \\
J131046.16+613521.8 & 2466 & 18.73 & 1.065 & 0.4569 & 1.093$\pm$0.110 & 0.921$\pm$0.120 & 0.216$\pm$0.084 & \nodata & B \\
J131204.26+494632.4 & 2334 & 17.61 & 1.135 & 0.6606 & 2.838$\pm$0.048 & 2.343$\pm$0.048 & 0.453$\pm$0.051 & 1.673$\pm$0.054 & B \\
J131405.08+041821.8 & 2172 & 18.14 & 0.943 & 0.6757 & 1.117$\pm$0.077 & 0.918$\pm$0.078 & 0.491$\pm$0.101 & 0.848$\pm$0.096 & A \\
J132746.16+484202.9 & 2334 & 18.20 & 1.029 & 0.5143 & 1.453$\pm$0.095 & 1.194$\pm$0.091 & 0.181$\pm$0.107 & 0.800$\pm$0.095 & B \\
J133719.31+594905.4 & 2452 & 17.80 & 0.658 & 0.5495 & 2.286$\pm$0.056 & 2.063$\pm$0.055 & 0.490$\pm$0.054 & 1.632$\pm$0.060 & A \\
J133819.96$-$025250.8 & 2172 & 18.73 & 0.578 & 0.5461 & 1.437$\pm$0.121 & 1.095$\pm$0.132 & 0.531$\pm$0.104 & 0.374$\pm$0.227 & B \\
J133924.99+622639.2 & 2466 & 18.85 & 1.149 & 0.6830 & 1.031$\pm$0.112 & 0.710$\pm$0.139 & 0.018$\pm$0.114 & 0.496$\pm$0.123 & B \\
J134050.87+642544.9 & 2466 & 18.42 & 1.139 & 0.5463 & 1.795$\pm$0.110 & 1.485$\pm$0.108 & 0.491$\pm$0.107 & 1.017$\pm$0.076 & D \\
J134541.67$-$020927.0 & 2172 & 19.21 & 1.247 & 0.6287 & 1.560$\pm$0.117 & 1.501$\pm$0.117 & 0.304$\pm$0.184 & 0.980$\pm$0.120 & B \\
J135741.65+052548.6 & 2178 & 18.78 & 0.740 & 0.6327 & 1.831$\pm$0.132 & 1.553$\pm$0.138 & 0.392$\pm$0.121 & 1.187$\pm$0.106 & A \\
J135805.08+393518.3 & 2172 & 17.85 & 0.879 & 0.6807 & 1.904$\pm$0.040 & 0.663$\pm$0.049 & 0.622$\pm$0.053 & 1.077$\pm$0.046 & A \\
J135805.08+393518.3 & \nodata & \nodata & \nodata & 0.6827 & 0.581$\pm$0.047 & 0.329$\pm$0.051 & 0.156$\pm$0.040 & 0.067$\pm$0.042 & B \\
J135939.34+572354.8 & 2452 & 18.19 & 1.134 & 0.4025 & 2.768$\pm$0.123 & 2.712$\pm$0.137 & 0.742$\pm$0.092 & \nodata & C \\
J140528.76+630325.6 & 2466 & 18.74 & 1.199 & 0.6837 & 1.670$\pm$0.092 & 1.473$\pm$0.098 & 0.278$\pm$0.095 & 1.391$\pm$0.097 & B \\
J140918.82+645521.5 & 2466 & 18.89 & 1.030 & 0.4312 & 1.545$\pm$0.139 & 1.142$\pm$0.152 & 0.058$\pm$0.114 & \nodata & C \\
J141515.47+011253.2 & 2172 & 18.92 & 1.038 & 0.5946 & 1.749$\pm$0.108 & 1.896$\pm$0.121 & 0.649$\pm$0.138 & 1.328$\pm$0.112 & B \\
J141607.28+552841.9 & 2452 & 18.41 & 0.772 & 0.5951 & 1.596$\pm$0.077 & 1.455$\pm$0.076 & 0.434$\pm$0.082 & 1.000$\pm$0.100 & B \\
J142021.45+594024.0 & 2452 & 18.90 & 1.251 & 0.6588 & 1.848$\pm$0.089 & 1.908$\pm$0.083 & 0.333$\pm$0.078 & 1.796$\pm$0.067 & A \\
J142119.42+611539.0 & 2466 & 18.12 & 0.656 & 0.6302 & 1.697$\pm$0.074 & 1.509$\pm$0.069 & 0.546$\pm$0.072 & 1.048$\pm$0.092 & B \\
J142609.74+594629.6 & 2452 & 17.85 & 0.743 & 0.5405 & 1.045$\pm$0.081 & 0.353$\pm$0.069 & 0.111$\pm$0.060 & 0.139$\pm$0.069 & B \\
J142816.70+585432.5 & 2452 & 17.85 & 0.780 & 0.3769 & 1.067$\pm$0.134 & 1.189$\pm$0.122 & 0.181$\pm$0.091 & \nodata & C \\
J143719.19+380437.7 & 2274 & 17.60 & 1.040 & 0.5792 & 1.430$\pm$0.059 & 1.219$\pm$0.062 & 0.121$\pm$0.076 & 0.480$\pm$0.106 & B \\
J144314.18$-$024722.1 & 2172 & 17.35 & 0.677 & 0.6504 & 2.860$\pm$0.031 & 2.593$\pm$0.035 & 0.728$\pm$0.036 & 2.036$\pm$0.037 & B \\
J145712.86+431849.9 & 2286 & 18.02 & 1.074 & 0.6314 & 1.253$\pm$0.079 & 0.904$\pm$0.077 & 0.352$\pm$0.093 & 0.454$\pm$0.073 & B \\
J150431.30+474151.2 & 2334 & 18.62 & 0.824 & 0.5942 & 1.082$\pm$0.100 & 0.788$\pm$0.092 & 0.133$\pm$0.089 & 0.139$\pm$0.092 & B \\
J151106.40+590111.0 & 2452 & 18.95 & 0.648 & 0.4642 & 1.755$\pm$0.122 & 1.323$\pm$0.146 & 0.222$\pm$0.106 & \nodata & B \\
J151413.59+553500.6 & 2452 & 18.58 & 1.325 & 0.6645 & 1.198$\pm$0.092 & 0.977$\pm$0.087 & 0.205$\pm$0.125 & 0.622$\pm$0.075 & B \\
J151505.12+041012.1 & 2172 & 18.52 & 1.272 & 0.5592 & 1.152$\pm$0.098 & 0.576$\pm$0.071 & 0.067$\pm$0.075 & 0.387$\pm$0.078 & A \\
J151528.47+454123.3 & 2334 & 18.75 & 0.800 & 0.3853 & 2.403$\pm$0.227 & 1.746$\pm$0.187 & 0.544$\pm$0.159 & \nodata & C \\
J152816.61+322258.0 & 2246 & 17.74 & 0.523 & 0.4923 & 1.255$\pm$0.064 & 0.688$\pm$0.062 & 0.031$\pm$0.046 & 1.218$\pm$0.105 & B \\
J153106.34+455750.6 & 2334 & 17.53 & 0.996 & 0.6757 & 1.252$\pm$0.063 & 0.833$\pm$0.052 & 0.052$\pm$0.051 & 0.680$\pm$0.069 & B \\
J153502.29+371324.6 & 2246 & 18.84 & 1.016 & 0.5834 & 1.390$\pm$0.164 & 1.164$\pm$0.142 & 0.308$\pm$0.116 & 0.197$\pm$0.117 & B \\
J154744.55+483603.4 & 2334 & 18.93 & 1.013 & 0.5212 & 1.092$\pm$0.129 & 1.201$\pm$0.132 & 0.218$\pm$0.107 & 0.478$\pm$0.143 & A \\
J155024.32+545338.0 & 2388 & 18.66 & 0.950 & 0.6664 & 1.100$\pm$0.098 & 0.608$\pm$0.096 & 0.136$\pm$0.094 & 0.238$\pm$0.117 & C \\
J155811.13+394549.9 & 2246 & 18.10 & 1.148 & 0.6582 & 1.055$\pm$0.074 & 0.819$\pm$0.061 & 0.243$\pm$0.066 & 0.586$\pm$0.058 & B \\
J161428.07+493004.4 & 2334 & 17.54 & 0.645 & 0.4256 & 1.972$\pm$0.057 & 1.916$\pm$0.057 & 0.510$\pm$0.061 & \nodata & A \\
J162806.69+434029.8 & 2286 & 17.74 & 1.021 & 0.6085 & 1.042$\pm$0.056 & 0.965$\pm$0.059 & 0.204$\pm$0.058 & 0.486$\pm$0.063 & A \\
J163842.86+204841.1 & 2222 & 17.35 & 1.054 & 0.6684 & 1.932$\pm$0.033 & 1.745$\pm$0.034 & 0.365$\pm$0.046 & 1.493$\pm$0.037 & A \\
J164346.90+434645.0 & 2286 & 17.74 & 0.939 & 0.5679 & 1.123$\pm$0.065 & 1.205$\pm$0.063 & 0.112$\pm$0.057 & 0.853$\pm$0.072 & B \\
J164544.69+375526.1 & 2246 & 18.22 & 0.602 & 0.5829 & 1.568$\pm$0.074 & 1.329$\pm$0.064 & 0.371$\pm$0.081 & 1.066$\pm$0.085 & B \\
J205601.68$-$001613.2 & 2172 & 18.05 & 0.521 & 0.4689 & 1.567$\pm$0.086 & 1.274$\pm$0.088 & 0.324$\pm$0.062 & \nodata & B \\
J211227.29+092012.2 & 2178 & 18.56 & 0.547 & 0.5190 & 1.134$\pm$0.091 & 1.208$\pm$0.088 & 0.403$\pm$0.062 & 0.973$\pm$0.136 & B \\
J211905.64+113924.4 & 2186 & 17.65 & 0.667 & 0.5429 & 1.665$\pm$0.072 & 0.953$\pm$0.065 & 0.300$\pm$0.058 & 0.370$\pm$0.063 & B \\
J220537.23+131629.1 & 2186 & 18.86 & 0.556 & 0.5203 & 1.492$\pm$0.134 & 1.702$\pm$0.136 & 0.353$\pm$0.106 & 0.877$\pm$0.164 & B \\
J233544.18+150118.3 & 2192 & 18.16 & 0.791 & 0.6801 & 1.215$\pm$0.060 & 1.262$\pm$0.056 & 0.556$\pm$0.072 & 1.065$\pm$0.059 & B \\
1629+120CAL & 2186 & 18.63 & 1.787 & 0.5313 & 1.400$\pm$0.070 & 1.350$\pm$0.070 & 0.310$\pm$0.080 & 0.710$\pm$0.100 & E \\
2328+0022CAL & 2172 & 17.83 & 1.306 & 0.6519 & 1.896$\pm$0.077 & 1.484$\pm$0.073 & 0.550$\pm$0.079 & 1.258$\pm$0.065 & E \\
2353$-$0028CAL & 2172 & 18.16 & 0.764 & 0.6044 & 1.601$\pm$0.082 & 1.292$\pm$0.083 & 0.606$\pm$0.080 & 1.024$\pm$0.104 & E \\

\hline

\end{tabular}
\tablenotetext{a}{Status Codes: A (high-probability DLA; see Table 2 and Figure 3), B (weak or absent absorption at the \MgII-predicted position of Ly$\alpha$ absorption; could possibly be a subDLA), C (spectrum cannot be used to determine whether a DLA is present), D (no flux visible near the \MgII-predicted position of Ly$\alpha$ absorption or at shorter wavelengths but an extremely rare, very high-$N_{\rm HI}$ DLA cannot be ruled out; see Table 2 and Figure 3), and E (DLA calibrator; see Table 2 and Figure 3).}
\end{table*}

The HST observations were taken between 9 August 2005 and 23 December 2006. Each quasar observation was allocated one HST orbit. The observing procedure was as follows. First we obtained two 150 sec direct images with ACS-HRC-F606W, with a small line dither (pointing maneuver) of 0.084 arcsec at a pattern orientation of 32.1 deg between them. The main purpose of the direct images was to use them to establish the wavelength calibrations on the prism images. Obtaining two direct images allowed most cosmic rays to be evaluated and removed in the individual images as well as in a combined drizzled image. Second, for the remainder of the orbit, we obtained two equally exposed prism images using the same dither pattern; individual exposure times for these observations ranged between 1050 sec and 1300 sec depending on the remaining available observation time in the orbit. The existence of two prism images also permitted the evaluation and removal of cosmic rays in the individual prism images and combined drizzled image. Of course, cosmic rays and other backgrounds were generally more problematic in the longer-exposure images, and in several instances we had to discard one of the prism images. As an example of the kinds of images that were obtained, the two prism images of a quasar that served as a DLA calibrator are shown in Figure 2.  

\begin{figure}
\vspace{0.0in}\centerline{
\includegraphics[width=0.95\columnwidth,clip,angle=0]{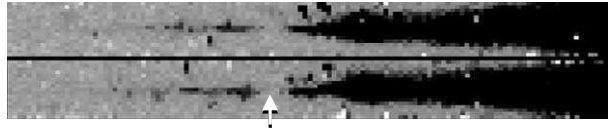}}
\vspace{0.0in}
\caption{The two dithered HST ACS-HRC-PR200L prism images for Q2353-0028 (top and bottom), which was one of our DLA calibration systems. The wavelength scale increases to the right. The quasar has a DLA marked by the white arrow at $z=0.604$ with $N_{\rm HI} = 3.5 \times 10^{21}$ atoms cm$^{-1}$ (RTN2006).}
\label{fig:calibrate}
\end{figure}

The data were processed using the aXe software developed by the ST-ECF team (K\"ummel et al. 2009). As noted above, the direct images were used to wavelength-calibrate the prism images. A one-dimensional prism spectrum was extracted from each of the two prism images, the spectra were flux calibrated, and then combined to form the final spectrum with an associated error array. 

An important part of evaluating the data involved examination of the individual two-dimensional prism images and the corresponding extracted one-dimensional prism spectra from the individual usable exposures. For example, as illustrated by Figure 2, the presence of a DLA is quite easy to observe in the prism images. However, the process of determining $N_{\rm HI}$ included some iteration because it is not possible to rely heavily on a guess for $N_{\rm HI}$  simply by making careful visual inspections, especially since the prism dispersion changes with wavelength. Thus, it was not sufficient to simply fit a DLA line to an extracted one-dimensional prism spectrum. We also examined the two-dimensional prism images to visually search for noise that may have influenced the original fit, and then adjusted the fit accordingly. 
In the end, a system was designated with status code A or B in Table 1 when its prism spectrum was judged to be of sufficient quality to permit a determination of whether a DLA line was present at the \MgII-predicted position of Ly$\alpha$ absorption. 
After these iterations, there were 35 status code A systems that were judged to have a significant probability (e.g., within 1$\sigma$ of $N_{\rm HI} = 1\times 10^{20}$ atoms cm$^{-2}$) of being DLAs given our estimate of $N_{\rm HI}$ and the associated error (see Section 3 and Table 2). The choice of using $N_{\rm HI} = 1\times 10^{20}$ atoms cm$^{-2}$ is meant to be a conservative choice (e.g., similar to that of Wolfe et al. 1986 when analyzing the ``Lick spectra'') in the sense that this list should not miss any classical DLAs with $N_{\rm HI} \ge 2\times10^{20}$ atoms cm$^{-2}$. Thus, our list will be complete for DLAs but not subDLAs. Note that the lowest $N_{\rm HI}$ reported in Table 2 is $8\times10^{19}$ atoms cm$^{-2}$, but it is $\sim2.5\sigma$ from having the classical DLA value. 

The 62 status code B systems were those judged to show no absorption or at best weak absorption at the \MgII-predicted position of Ly$\alpha$, but with insufficient strength to be a classical DLA; thus the status code B systems may or may not be subDLAs, but they can be categorized as non-DLAs in any statistical study. 

The 11 status code C systems in Table 1 were those with unusable spectra; in one case (J155024.32+545338.0) the system lies within 3300 km s$^{-1}$ of Ly$\beta$$+$\OVI\ emission and cannot be resolved, in five cases the post-observation results indicated that the redshifts were in fact too low ($z<0.42$) to be usefully observed with the prism (i.e., the prism's sensitivity was too low at shorter wavelengths), and in the remaining cases the spectra had insufficient quality (low signal) at the \MgII-predicted position of Ly$\alpha$ absorption to make any determination. 

One object in Table 1 (J134050.87+642544.9) was designated as status code D because it has no visible flux within 60 \AA\ (observed frame) of the \MgII-predicted postion of Ly$\alpha$ absorption or at shorter wavelengths, and therefore it could be an extremely rare, very high-$N_{\rm HI}$ DLA. We checked the SDSS spectrum of this object to search for a \MgII\ absorption system that might give rise to Lyman limit absorption and explain the lack of UV flux, but found none. 

At the end of Table 1 we present information on the three observed DLA calibrators and designate them as status code E.

After estimating the continuum levels across the predicted locations of Ly$\alpha$ absorption using a simple straight-line fit, the spectra were then continuum-normalized at these locations.\footnote{The fitted-continuum levels occasionally had mild slopes and, of course, fitting a continuum was not reliable for status code C systems.} This normalization was done for the 35 status code A systems in 34 quasar spectra, the one status code D system and the three status code E DLA calibrator systems, and the results are shown in Figure 3. (This was also done for the 62 status code B systems, but the results are not shown.) Of course, using this procedure means that other portions of the spectra not near the predicted locations of Ly$\alpha$ absorption were not normalized near their local continuum levels. For example, strong broad emission lines were generally not fitted with a pseudo-continuum, so they generally appear above the unit continuum levels and blends of strong Ly$\alpha$ forest absorption (but non-DLAs) will appear as unresolved complexes of absorption depressing the unit continuum levels; the overall spectrum shapes may also not be flat. As described in the the next section, the Ly$\alpha$ absorption lines were then fitted with Voigt damping profiles in order to estimate the $N_{\rm HI}$ values and associated errors. 

\begin{figure*}
\centerline{
\includegraphics[width=2.2\columnwidth,angle=0]{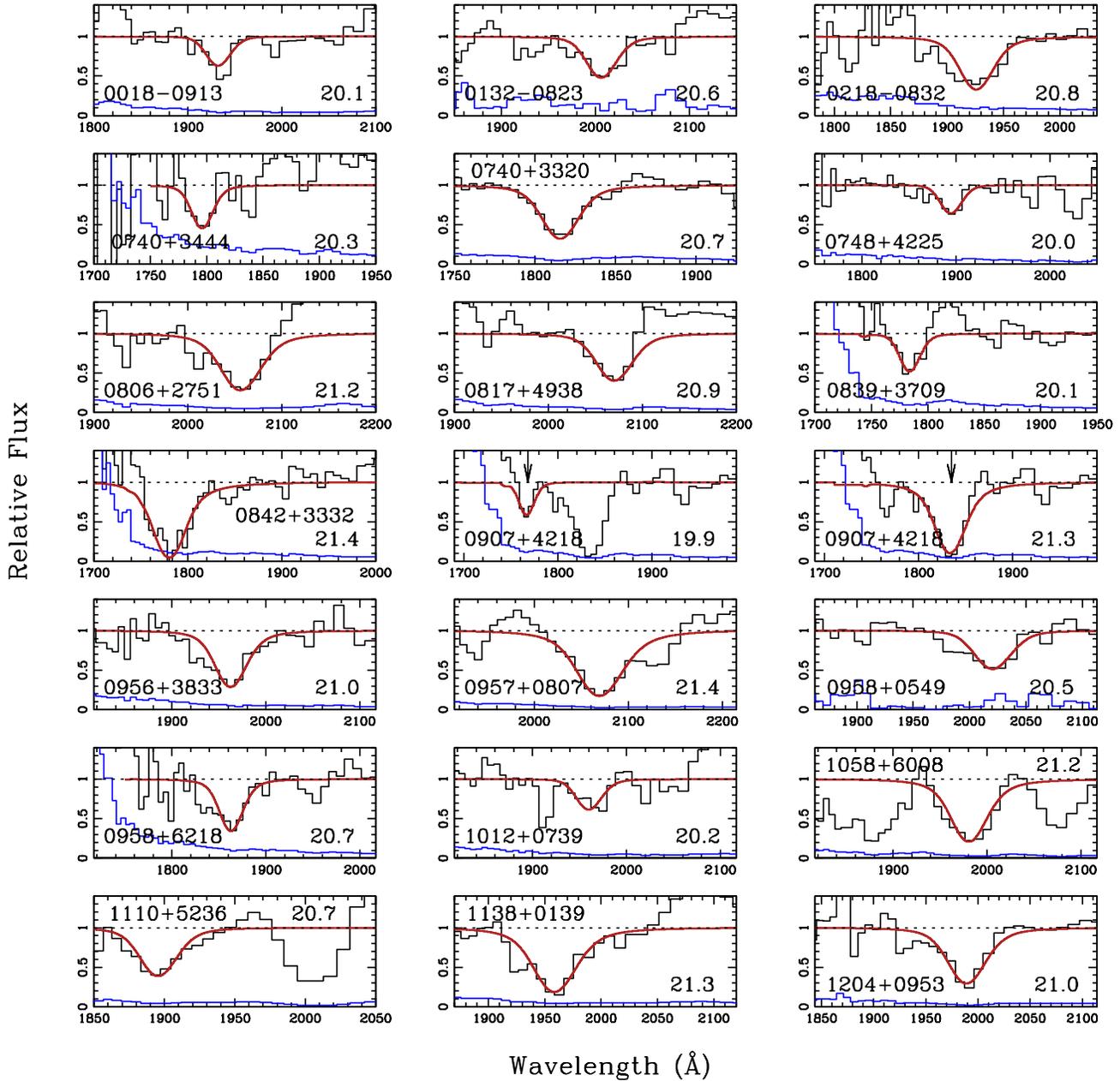}}
\caption{The HST UV prism spectra of the Ly$\alpha$ regions in 35 strong-\MgII\ systems which have a significant probability of being DLAs (0907+4218 has two systems), plus one additional strong-\MgII\ system that may be an extremely high column density DLA (1340+6425). The last three spectra are DLA calibrators. Shown are the relative fluxes of the prism spectra normalized near Ly$\alpha$, the errors in normalized fluxes (in blue, see online version) derived using the aXe software reduction package or the empirical error (whichever is worse), and the fits to the DLAs (red, online version) for the $N_{\rm HI}$ values reported in Table 2. The shown errors in normalized fluxes were used to derive the $N_{\rm HI}$ errors reported in Table 2. The J2000 coordinate names and fitted log($N_{\rm HI}$) values are labeled on the prism spectrum panels. All the strong-\MgII\ systems have $z_{abs}=[0.42,0.70]$ as reported in Tables 2 and 3.}
\label{fig:specs}
\end{figure*}


\begin{figure*}

\vspace{0.0in}\centerline{
\includegraphics[width=2.2\columnwidth,angle=0]{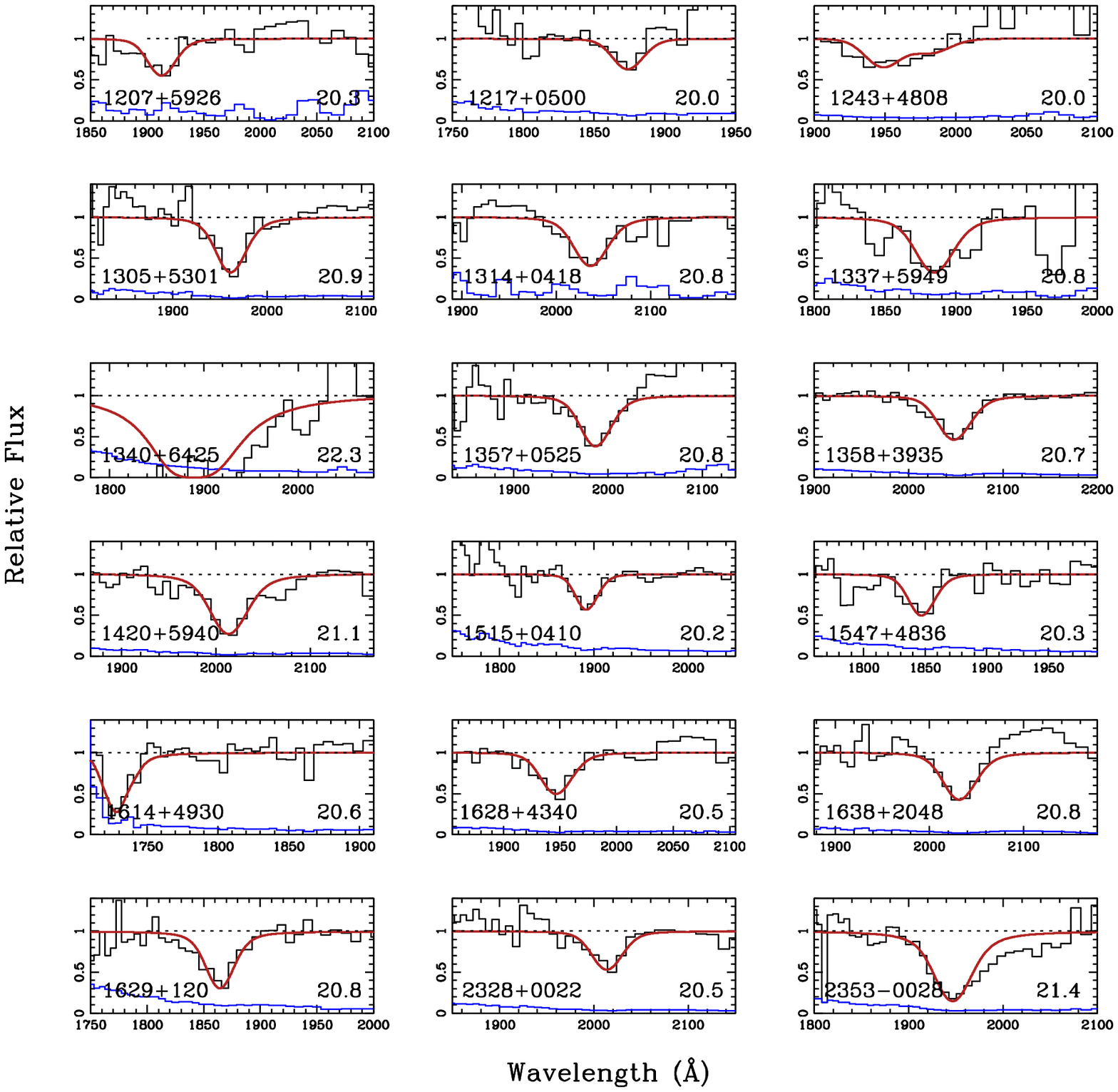}}
\addtocounter{figure}{-1}
\caption{Continued from the previous page: the HST UV prism spectra of the Ly$\alpha$ regions in 35 strong-\MgII\ systems which have a significant probability of being DLAs (0907+4218 has two systems), plus one additional strong-\MgII\ system that may be an extremely high column density DLA (1340+6425). The last three spectra are DLA calibrators. Shown are the relative fluxes of the prism spectra normalized near Ly$\alpha$, the errors in normalized fluxes (in blue, see online version) derived using the aXe software reduction package or the empirical error (whichever is worse), and the fits to the DLAs (red, online version) for the $N_{\rm HI}$ values reported in Table 2. The shown errors in normalized fluxes were used to derive the $N_{\rm HI}$ errors reported in Table 2.  The J2000 coordinate names and fitted $\log (N_{\rm HI}$) values are labeled on the prism spectrum panels. All the strong-\MgII\ systems have $z_{abs}=[0.42,0.70]$ as reported in Tables 2 and 3.}
\label{fig:specs}
\end{figure*}
\section{Neutral Hydrogen Column Density Results}

\begin{table*}
\small
\caption{Column density measurements}
\begin{tabular}{lccc}
\hline
\hline
Quasar  & \MgII\ $z_{abs}$ & $\rm  N_{\rm HI}$  & Notes \\
 & & atoms cm$^{-2}$ \\
\hline
J001855.22$-$091351.1 & 0.5838 &  $(1.3\pm0.3) \times 10^{20}$ &\\
J013209.75$-$082349.9 & 0.6467 &  $(4.0\pm1.1) \times 10^{20}$ &\\
J021820.11$-$083259.4 & 0.5899 &  $(7.0\pm2.0) \times 10^{20}$ &\\
J074043.44+344407.7 & 0.4783 &  $(2.0\pm1.1) \times 10^{20}$ &\\
J074054.05+332006.5 & 0.4917 &  $(4.7\pm1.2) \times 10^{20}$ &\\
J074816.97+422509.3 & 0.5579 &  $(1.0\pm0.4) \times 10^{20}$ &\\
J080601.80+275108.2 & 0.6922 &  $(1.5\pm0.2) \times 10^{21}$ &\\
J081748.20+493821.4 & 0.6855 &  $(7.5\pm1.2) \times 10^{20}$ &\\
J083900.67+370901.4 & 0.4697 &  $(1.3\pm0.6) \times 10^{20}$ &\\
J084200.76+333214.8 & 0.4750 &  $(2.8\pm0.8) \times 10^{21}$ &\\
J090757.59+421823.6 & 0.4540 &  $(8.0\pm4.5) \times 10^{19}$ &\\
J090757.59+421823.6 & 0.5106 &  $(2.0\pm0.5) \times 10^{21}$ &\\
J095648.58+383339.0 & 0.6196 &  $(1.0\pm0.1) \times 10^{21}$ &\\
J095740.06+080732.2 & 0.6975 &  $(2.8\pm0.3) \times 10^{21}$ &\\
J095844.07+054941.0 & 0.6557 &  $(3.5\pm1.2) \times 10^{20}$ &\\
J095848.10+621839.0 & 0.5357 &  $(5.0\pm2.0) \times 10^{20}$ &\\
J101211.11+073949.9 & 0.6164 &  $(1.5\pm0.4) \times 10^{20}$ &\\
J105823.13+600805.6 & 0.6436 &  $(1.7\pm0.3) \times 10^{21}$ & a \\
J111013.80+523607.1 & 0.5561 &  $(4.5\pm0.5) \times 10^{20}$ & b\\
J113823.71+013924.8 & 0.6130 &  $(1.8\pm0.4) \times 10^{21}$ &\\
J120449.73+095335.2 & 0.6401 &  $(1.1\pm0.2) \times 10^{21}$ &\\
J120743.71+592648.2 & 0.5765 &  $(1.9\pm0.6) \times 10^{20}$ &\\
J121753.03+050030.8 & 0.5413 &  $(1.0\pm0.4) \times 10^{20}$ &\\
J124312.52+480838.4 & 0.6053 &  $(1.1\pm0.3) \times 10^{20}$ &\\
J130544.31+530136.2 & 0.6148 &  $(8.0\pm2.0) \times 10^{20}$ &\\
J131405.08+041821.8 & 0.6757 &  $(6.8\pm1.5) \times 10^{20}$ &\\
J133719.31+594905.4 & 0.5495 &  $(6.0\pm1.0) \times 10^{20}$ &\\
J134050.87+642544.9 & 0.5463 & ($\gtrsim 2 \times 10^{22}$) & c \\
J135741.65+052548.6 & 0.6327 &  $(6.5\pm0.7) \times 10^{20}$ &\\
J135805.08+393518.3 & 0.6807 &  $(5.0\pm1.0) \times 10^{20}$ & \\
J142021.45+594024.0 & 0.6588 &  $(1.4\pm0.2) \times 10^{21}$ &\\
J151505.12+041012.1 & 0.5592 &  $(1.6\pm0.7) \times 10^{20}$ &\\
J154744.55+483603.4 & 0.5212 &  $(2.0\pm0.5) \times 10^{20}$ &\\
J161428.07+493004.4 & 0.4256 &  $(4.0\pm2.0) \times 10^{20}$ &\\
J162806.69+434029.8 & 0.6085 &  $(3.0\pm0.8) \times 10^{20}$ &\\
J163842.86+204841.1 & 0.6684 &  $(6.0\pm1.0) \times 10^{20}$ &\\
1629+120CAL              & 0.5313 & $(5.8\pm 1.1) \times 10^{20}$ & d \\
2328+0022CAL            & 0.6519 & $(3.2\pm 0.7) \times 10^{20}$ & e \\
2353$-$0028CAL         & 0.6044 & $(2.3\pm 0.4) \times 10^{21}$ & f \\

\hline

\end{tabular}
\tablenotetext{a}{J1058+6008 is a broad absorption line QSO, but no strong BAL features are expected at the location of the DLA.} 

\tablenotetext{b}{J1110+5236 has a broad absorption feature centred near 2010 \AA, which is part of a BAL system.} 

\tablenotetext{c}{The very high $N_{\rm HI}$ constraint for J134050.87+642544.9 is based on the possibility that the prism spectrum flux disappears at short wavelength due to a strong DLA. However, such systems are expected to be extremely rare. See the discussion in \S3 of the text.} 

\tablenotetext{d}{Rao et al. (2003) report $N_{\rm HI} =  (5.0\pm 1.0) \times 10^{20}$ atoms cm$^{-2}$ based on HST-STIS spectroscopy.}

\tablenotetext{e}{RTN2006 report $N_{\rm HI} =  (2.1\pm 0.3) \times 10^{20}$ atoms cm$^{-2}$ based on HST-STIS spectroscopy.}

\tablenotetext{f}{RTN2006 report $N_{\rm HI} =  (3.5\pm 1.2) \times 10^{21}$ atoms cm$^{-2}$ based on HST-STIS spectroscopy.} 

\end{table*}

The 38 continuum-normalized prism spectra, f$_{\lambda}$, and associated error arrays, $\sigma_{f_{\lambda}}$ (derived using the aXe software), are shown in Figure 3. They were used to measure Ly$\alpha$ in 39 strong-\MgII\ systems. Two of the systems are present in the prism spectrum of J090757.59+421823.6. The 
first 36 systems in Figure 3 (in right ascension order) are those from our new sample which were found to have a significant (35 with status code A) or interesting (one with status code D) probability of being DLAs. The last three spectra in Figure 3 are of the DLA calibrator quasars (status code E).  

Establishing the continuua levels in low-resolution DLA survey spectra has, unfortunately, always been somewhat of an art. With a low-resolution spectrum there is the problem of unresolved absorption blends of Ly$\alpha$ forest lines. An additional complication is the non-linear dispersion (resolution) in a slitless prism spectrum, which results in an asymmetric DLA profile (see Figure 1).  Fortunately, the \HI\ column density is uniquely determined by the absorption rest equivalent width and, for a given continuum, a Voigt profile �fit� convolved with the instrument dispersion solution gives a unique value for the column density. Thus, setting the continuum level properly is important, and we have carefully considered the various relevant issues.

Continuum placement was a matter of identifying emission-line free regions on both sides (or sometimes only one side) of the predicted location of any DLA line, and then judging where the continuum was at the relevant wavelength. Given the varying properties of the background quasar spectra, the procedure necessarily changed from absorption system to absorption system. Note that, in some cases, the regions used to set the continuum level may not be included in the plots. As an extra precaution, the first three authors of this paper thoroughly examined the possibly ambiguous cases by looking at prism images and extracted spectra (before addition) to evaluate whether a continuum placement was reasonable. As an example, it's worth noting that the apparent excess emission to the red of the DLA in J083900.67+370901.4 is in fact real broad emission due to Ly$\beta$$+$\OVI\ in the quasar, and not an artifact of continuum placement. Another example is in J084200.76+333214.8; its spectrum becomes noisy on the short wavelength end of the spectrum, and that part was not used to set the continuum.

To determine $N_{\rm HI}$ values, theoretical Ly$\alpha$ Voigt profiles were shifted to the \MgII\ absorption redshifts and convolved to the prism resolution using the aXe task ``simdata.'' The resulting simulated spectra were then 
adjusted to match the continuua slopes of the extracted prism spectra and overlaid to determine (by eye) the best-fitting $N_{\rm HI}$ values. The Voigt profiles that were judged to produce 
the best fits to the observed damped profiles are shown in Figure 3 as solid red lines (online version) relative to the unit continuum levels shown as black 
dashed lines near the DLAs. 

We were able to check the validity of the aXe derivations of $\sigma_{f_{\lambda}}$ for the vast majority of the spectra in Figure 3.\footnote{In six of the 38 cases we 
discarded one of the two extracted prism spectra because it appeared to be of much poorer quality than the other.} We did this by comparing the two individual 
f$_{\lambda}$ spectra to derive an empirical error array, $\sigma^{emp}_{f_{\lambda}}$. In most cases $\sigma_{f_{\lambda}}$ and $\sigma^{emp}_{f_{\lambda}}$ were 
found to be consistent, but in some cases they were not. As discussed below, when there was a significant discrepancy, this led us to use a larger continuum 
placement error to estimate the error in $N_{\rm HI}$.  

To determine the error in $N_{\rm HI}$ a spectrum was re-normalized by dividing by 1$\pm$$\sigma_{f_{\lambda}}$ or 1$\pm$$\sigma^{emp}_{f_{\lambda}}$ (whichever 
was larger), and the damped profiles were re-fit with Voigt profiles to determine the likely range in $N_{\rm HI}$. Generally the $\pm$$\sigma_{N_{\rm HI}}$ errors were 
not symmetric, so the adopted value for $\sigma_{N_{\rm HI}}$ was taken to be the one with the largest absolute value. 
Finally, in some cases tests indicated that resulting errors in $N_{\rm HI}$ for lower $N_{\rm HI}$ systems were unrealistically low, and we concluded that we 
should never report the error in $N_{\rm HI}$ to be $<$20 per cent for a system with $N_{\rm HI} < 10^{21}$ atoms cm$^{-2}$. Therefore, in those cases we report a 20 per cent error. 
Our derivations of $N_{\rm HI}$ using the prism spectra of the three DLA calibrators (see the last three entries in Table 2 and their footnotes) confirm that the 
algorithm we used to report the $\sigma_{N_{\rm HI}}$ values is reasonable and consistent with known results. Note that the algorithmic $N_{HI}$ error we report for 2353$-$0028CAL is smaller than the error reported in RTN2006. Also, it should be mentioned that some of our determined $N_{\rm HI}$ percentage errors are smaller than typical $N_{\rm HI}$ errors derived from analyses of SDSS spectra (e.g., Noterdaeme et al. 2009). This may seem surprising given the differences in data quality and resolution in these two very different spectral data sets. However, it is also the case that Ly$\alpha$ forest absorption is both more frequent and problematic at the higher redshifts probed by the SDSS spectra. 
 
The final results are given in Table 2. The redshifts in Table 2 are those measured from \MgII\ absorption in the SDSS spectra, since measuring a redshift from a damped absorption 
line in a prism spectrum is not as accurate. The $N_{\rm HI}$ value for J134050.87+642544.9 (the one status code D absorption system) is taken to be at least $N_{\rm HI} \sim 2\times10^{22}$ atoms cm$^{-2}$ based on the assumption that only the long-wavelength side of the damping profile is visible in the prism spectrum;  
however, this $N_{\rm HI}$ value is so large that it is unlikely to be present in a sample of our size, so this might not in fact be a DLA. We note that in the large high-redshift DLA 
samples identified by Noterdaeme et al. (2012, 2014) using SDSS spectra, only one in a thousand had $N_{\rm HI} > 10^{22}$ atoms cm$^{-2}$. 

\section{Significance}

The main results of our study are presented in Table 2, namely the identification of 36 new, high-$N_{\rm HI}$ systems that have a significant or interesting probability 
of being DLAs at $z=[0.42,0.70]$, which is a relatively narrow low-redshift cosmic time interval corresponding to $t\approx[7.2,9.0]$ Gyrs after the Big Bang. 
While some cosmic evolution might be expected over this small interval, any evolutionary effects and observational biases would be minimal in comparison 
to the alternative of working over the entire redshift interval covered by UV spectra, i.e., $z<1.65$, which corresponds to the most recent $\sim 9.6$ Gyrs of the history of the Universe. 

Given the availability of the sample presented here, 
we suggest that observational biases and possible evolutionary effects could be more easily treated if future observational 
studies of low-redshift DLAs were concentrated within the $z=[0.42,0.7]$ redshift interval.\footnote{Nine additional DLAs could be added to this sample by including previously known DLAs at $z=[0.42,0.70].$} 
For comparison, the very large sample of high-redshift DLAs identified by Noterdaeme et al. (2012) 
at $z=[2.0,4.0]$ corresponds to a cosmic time interval of $\sim 1.7$ Gyr, which is $t\approx[1.5,3.2]$ Gyrs after the Big Bang. Noterdaeme et al. (2012) do find clear evidence for cosmic evolution, but cosmic evolution of DLAs appears to be more rapid at higher redshifts.   

It is important to note that, given that the $N_{\rm HI}$ values were derived using low-resolution prism spectra, the present work would clearly benefit from improved UV spectroscopy to more accurately determine the $N_{\rm HI}$ values. Such results would, of course, be beneficial to any follow-up work.

In any case, this new list of ``DLAs'' at $z=[0.42,0.70]$ is valuable for future follow-up studies of the neutral gas component of the Universe (e.g., metallicities, dust, molecular 
fractions, star formation, kinematics, associated galaxies, and clustering environment). The low redshifts and size 
of the sample makes it unique
for the purpose of identifying associated ``DLA galaxies,'' which has proved to be very difficult and sporadic at high redshift. Truly comprehensive studies of 
this sample would provide a framework for describing DLAs and their connections to galaxies that theory would have to explain. We hope that observers 
pursue this challenge. 

\section*{Acknowledgments}

These results are based on data obtained from the Sloan Digital Sky Survey (SDSS) and on observations made with the Hubble Space Telescope (HST) operated by STScI-AURA for NASA/ESA. We thank Martin K\"ummel for his advice on using the aXe software.  This work was supported from a grant from the Space Telescope Science Institute.  RH was supported by an undergraduate summer research award from the Brockport Physics Department. We thank an anonymous referee for suggesting that we clarify parts of the paper. 

The SDSS is managed by the Astrophysical Research Consortium for the Participating Institutions. The Participating Institutions are the American Museum of Natural History, Astrophysical Institute Potsdam, University of Basel, University of Cambridge, Case Western Reserve University, University of Chicago, Drexel University, Fermilab, the Institute for Advanced Study, the Japan Participation Group, Johns Hopkins University, the Joint Institute for Nuclear Astrophysics, the Kavli Institute for Particle Astrophysics and Cosmology, the Korean Scientist Group, the Chinese Academy of Sciences (LAMOST), Los Alamos National Laboratory, the Max-Planck-Institute for Astronomy (MPIA), the Max-Planck-Institute for Astrophysics (MPA), New Mexico State University, Ohio State University, University of Pittsburgh, University of Portsmouth, Princeton University, the United States Naval Observatory, and the University of Washington. 

{}

\label{lastpage}
\end{document}